\documentclass[journal=jacsat,manuscript=article]{achemso}

\usepackage[version=3]{mhchem}
\usepackage{amssymb}
\usepackage{graphicx}
\usepackage{dcolumn}
\usepackage{bm}
\usepackage[normalem]{ulem}
\usepackage{xcolor}
\usepackage{soul}
\usepackage{subcaption}

\newcommand{\bulk}{\mathrm{bulk}}
\newcommand{\conf}{\mathrm{conf}}
\newcommand{\linear}{\mathrm{linear}}
\newcommand{\ring}{\mathrm{ring}}
\newcommand{\micelle}{\mathrm{micelle}}

\title{Polymer Topology and the Depletion Interaction}

\author{Mauro L Mugnai}
\email{mm4994@georgetown.edu}
\affiliation{
 Institute for Soft Matter Synthesis and Metrology, Georgetown University, Washington, DC, 20057, USA\\
}

\begin{document}


\begin{abstract}
Using a theoretical model we show that ideal ring polymers are stronger depletants than ideal linear polymers of equal radii of gyration, but not of equal hydrodynamic radii. The difference in the depletion-induced force profile is largely controlled by the thickness of the depletion layer. Theory suggests that this thickness is equal to the average extent of a polymer along the direction perpendicular to the surfaces of the colloids. Within the limits of finite-size effects, Molecular Dynamics simulations support this conclusion.  
\end{abstract}


In 1954, Asakura and Oosawa (AO) developed a theory that showed that colloids experience an attractive interaction when immersed in a dilute solution of non-adsorbing macromolecules~\cite{Asakura1954JChemPhys}, also referred to as ``depletants''~\cite{Miyazaki2022JCP}. This attraction is due to the increase in depletant entropy occurring when colloids come into contact and increase the volume accessible to the macromolecules.
The magnitude of the depletion-induced attractive force between colloids is given by the osmotic pressure of the depletants, and the range is controlled by the size and shape of the macromolecules~\cite{Asakura1954JChemPhys,Asakura1958JPolSci,Piech2000JCIS,Briscoe2015COCIS,Kamien1999PRE,Lim2016SoftMatter,Denton2021JCP,LekkerkerkerBook}. 
Thus, the depletion interaction can be tuned by modifying the size and concentration of macromolecules without changing their chemistry or surface properties of the colloids~\cite{Likos2001PhsRep}.
Various experiments monitoring the interaction between colloids in dilute solutions of different depletants have confirmed theoretical predictions.~\cite{Evans1988Macromol,Richetti1992PRL,Kuhl1996Langmuir,Rudhardt1998PRL,Verma1998PRL,Crocker1999PRL,Sheffold2024NatComm}
Colloidal dispersion can be destabilized by adding depletants~\cite{Vrij1976PureApplChem} leading to phase diagrams which depend on the ratio of colloid and polymer size.~\cite{Lekkerkerker1992EPL,Calderon1993EPL,Ilett1995PRE,Dijkstra1999JoPCM} Depletion forces can be tuned to control self-assembly of colloidal particles~\cite{Sacanna2010Nature,Manna2010NanoLett} and aggregation of bacteria~\cite{Biggs2005Langmuir,Boluk2012ChemEngJ}, moreover they can lead to polymer collapse and stabilize the folded state of a protein.~\cite{Shaw1991PRA,Cheung2005PNAS} 

Recently, following a theoretical/computational prediction by Chubak {\it et al}~\cite{Chubak2018MolPhys}, Moghimi {\it et al}~\cite{Moghimi2024PhysRevRes} measured the rheology a of colloidal gel stabilized via depletion interaction induced by adding to the solution either linear or ring polymers. They found that, at given hydrodynamic radius of the depletant and at given amount of polymer, the storage modulus of the colloidal gel was larger in the presence of ring polymers, and that a lower concentration of ring polymer was sufficient to observe the onset of gelation. Thus, they concluded that ring polymers are stronger depletants than linear polymers of equal size. The authors further substantiated this observation by deploying a multi-scale coarse-graining approach for the solution of polymers and colloids, which shows that ring polymers give rise to stronger depletion interaction as compared to linear polymers of identical radius of gyration~\cite{Moghimi2024PhysRevRes}.

Here, we investigate why the change in polymer topology, from linear to ring, affects the depletion interaction. Polymers can be ``squished'' between colloidal particles. However, the confinement reduces the number of conformations that they can populate and thus polymers are entropically repelled from the space between colloids. If a ring polymer is a stronger depletant, it is reasonable to hypothesize that confinement between colloids causes a sharper reduction in entropy for ring polymers than for linear polymers of equal size. To test this idea, it is necessary to develop a theoretical model that can account for all the allowed polymer conformations and how the confinement impacts depletion forces. 
Here, we combine existing theoretical results for ideal linear~\cite{Asakura1954JChemPhys,Casassa1967JPolSci,Edwards1969JPhysA,Richmond1974ChemPhysLett,Gorbunov1984JPolSciUSSR,DoiBook,deGennesBook,LekkerkerkerBook} and ring~\cite{Gorbunov1984JPolSciUSSR} polymers and we find that at a given radius of gyration, but not hydrodynamic radius, ring polymers give rise to stronger depletion interaction. Building on known results~\cite{Tuinier2000JCP,LekkerkerkerBook}, we show that the strength of depletion interaction is controlled by the thickness of depletion layer around the colloids, $\xi$, and we that $2\xi$ is equal to the extent of the polymer along one direction averaged over all conformations, $\langle \Delta \rangle$. Within finite size effects, numerical results support this conclusion. 
\begin{figure}
\includegraphics[trim={0 4cm 0 0},width=0.5\columnwidth]{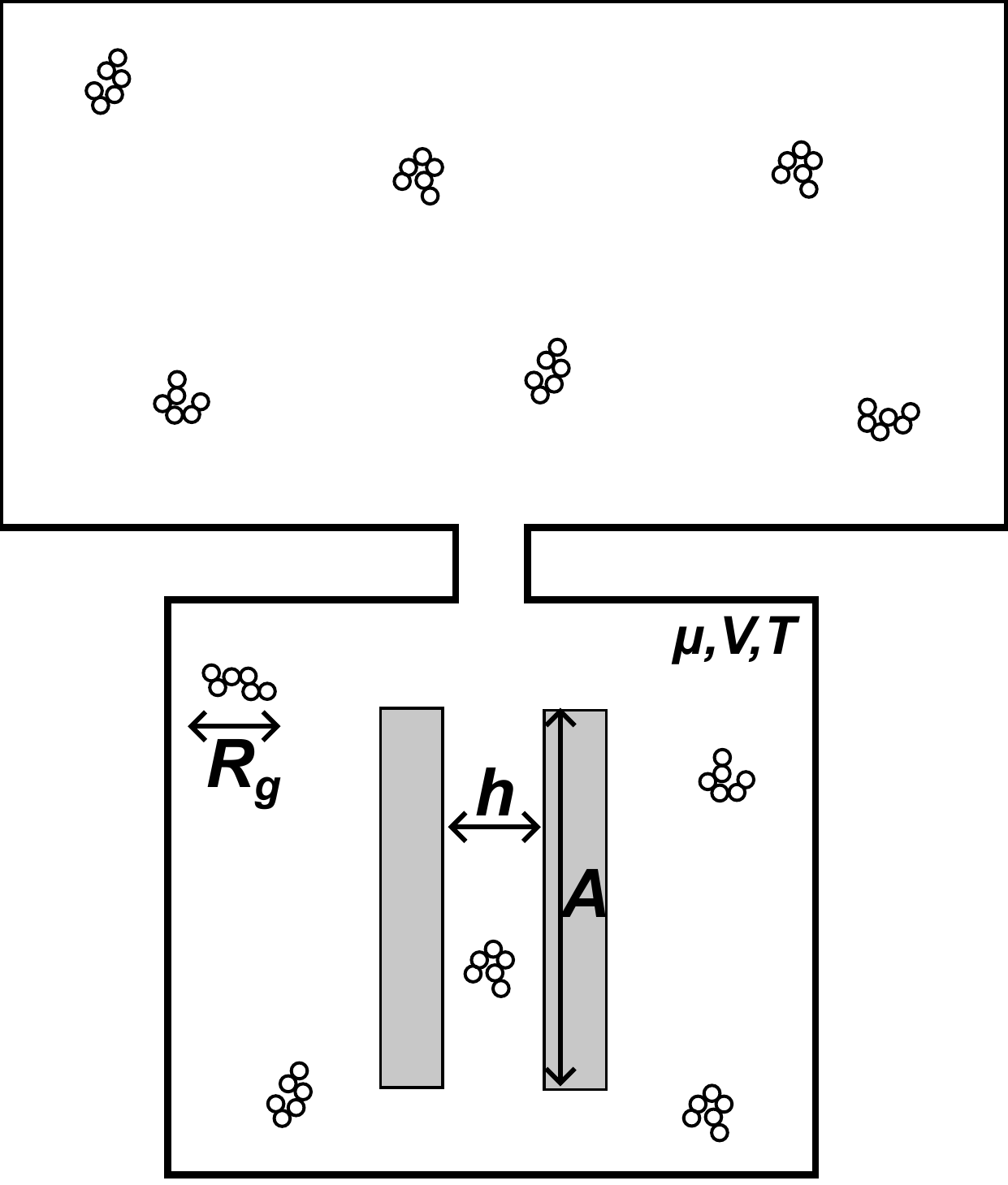}
\caption{\label{Fig:Model} Pictorial representation of the model.
Two colloidal plates (gray) of area $A$ and polymers (beads) of radius of gyration $R_g^2\ll A$ are immerse a chamber of volume $V$. A thermostat and a polymer reservoir ensure that the temperature of the chamber is $T$, and the chemical potential of the polymers is equal to $\mu$. }
\end{figure}

Consider a solution of polymers enclosed in a container of volume $V$ and in contact with a thermostat kept at temperature $T$ and with polymer reservoir at chemical potential $\mu$ (see Fig.~\ref{Fig:Model}). Let the solution also feature two colloidal plates of area $A \gg R_g^2$, where $R_g^2 = 1/(2n^2)\sum_{i=1}^n\sum_{j=1}^n \langle r_{ij}^2 \rangle$ is the square of the radius of gyration of the polymers of length $n$, $r_{ij}$ is distance between two monomers, and $\langle \cdots \rangle$ indicates an ensemble average~\cite{DoiBook}. We imagine the plates to be parallel to each other and at a surface-to-surface distance equal to $h$. The polymers are dilute, so that they do not interact with each other and can be treated as an ideal gas. In addition, the polymers are assumed to be repelled by the colloids so that they cannot be adsorbed onto their surfaces.
As a result, in addition to their hard-core interaction, the colloidal plates experience the following polymer-induce force ($f_p$) and potential energy ($V_p$)~\cite{LekkerkerkerBook},
\begin{equation}
\begin{aligned}
f_p(h) &= -\Pi A\Big\{1-\frac{\partial}{\partial h}\Big[h\frac{Z_\conf(hA)}{Z_\bulk(hA)}\Big]\Big\},\\
V_p(h) &= \int_h^\infty dh' f_p(h')
\end{aligned}
\label{Eq:fp}
\end{equation}
where $k_B$ is the Boltzmann constant, $T$ is temperature, $\Pi=k_BT\rho_{bulk}$ is the osmotic pressure of the macromolecules whose bulk density is $\rho_{bulk}$. Furthermore, $Z_\bulk(hA)$ and $Z_\conf(hA)$ are configurational partition functions constructed by placing the first monomer of the polymer in a volume $hA$ and either leaving the remaining ones unconstrained [$Z_\bulk(hA)$] or by confining them along the $x$-axis between repulsive walls separated by a distance $h$ [$Z_\conf(hA)$]. The logarithm of the ratio between these partition functions is the free energy penalty for confining the polymer. The potential energy in Eq.~\ref{Eq:fp} is defined as the work necessary to bring to infinite distance two plates initially separated by $h$. 

Polymer topology enters Eq.~\ref{Eq:fp} through the ratio of partition functions. Let the polymers be ideal chains, in which consecutive monomers are joined by a spring of mean-squared elongation equal to $b$ and stiffness $3k_BT/(2b^2)$, so that the energy of interaction between bonded monomers $i$ and $j$ is $3k_BT/(2b^2)r^2_{ij}$. We ignore volume exclusion.
Under these approximations, the three Cartesian components of the partition function are separable, and we can effectively solve a one-dimensional problem.
Using well known results for polymers under confinement~\cite{Asakura1954JChemPhys,Casassa1967JPolSci,Edwards1969JPhysA,Richmond1974ChemPhysLett,DoiBook,deGennesBook,LekkerkerkerBook}, we obtain the following for a linear polymer,
\begin{equation}
\begin{aligned}
\frac{Z_\conf^\linear(h)}{Z_\bulk^\linear(h)} &= \frac{8}{\pi^2}\sum_{p\in\text{odd}}\frac{1}{p^2}e^{-\pi^2(\frac{R_g}{h})^{2}p^2},
\end{aligned}
\label{Eq:linear}
\end{equation}
where $R_g = b(n/6)^{1/2}$ when the number of monomers in the polymer $n\gg 1$~\cite{DoiBook}. 
Similarly, for a ring polymer the result is,~\cite{Gorbunov1984JPolSciUSSR}
\begin{equation}
\begin{aligned}
\frac{Z_\conf^\ring(h)}{Z_\bulk^\ring(h)} &= \sqrt{8\pi}\frac{R_g}{h}\sum_{p=1}^\infty e^{-2\pi^2(\frac{R_g}{h})^2p^2}, 
\end{aligned}
\label{Eq:ring}
\end{equation}
where $R_g = b(n/12)^{1/2}$ for $n\gg 1$~\cite{Zimm1949JChemPhys}. Finally, we model a micelle depletant as a hard sphere of radius $R$ and radius of gyration $R_g= \sqrt{3/5}R$. The micelle can be inserted in the space between colloids only if its diameter is smaller than $h$, and its center can be moved only by $h-2R$ between the walls. 
Overall, we obtain,
\begin{equation}
\begin{aligned}
\frac{Z_\mathrm{conf}^\micelle(h)}{Z_\mathrm{bulk}^\micelle(h)} = \Big(1-\frac{2\sqrt{\frac{5}{3}}R_g}{h} \Big)\Theta\Big(h-2\sqrt{\frac{5}{3}}R_g\Big),
\end{aligned}
\label{Eq:micelle}
\end{equation}
where the step-function $\Theta(x)=1$ for $x>0$ and $=0$ otherwise.
Note that in all of these cases $Z_\conf/Z_\bulk$ is controlled by $h/R_g$.

\begin{figure*}
\begin{subfigure}[b]{0.45\textwidth}
\includegraphics[width=1.0\columnwidth]{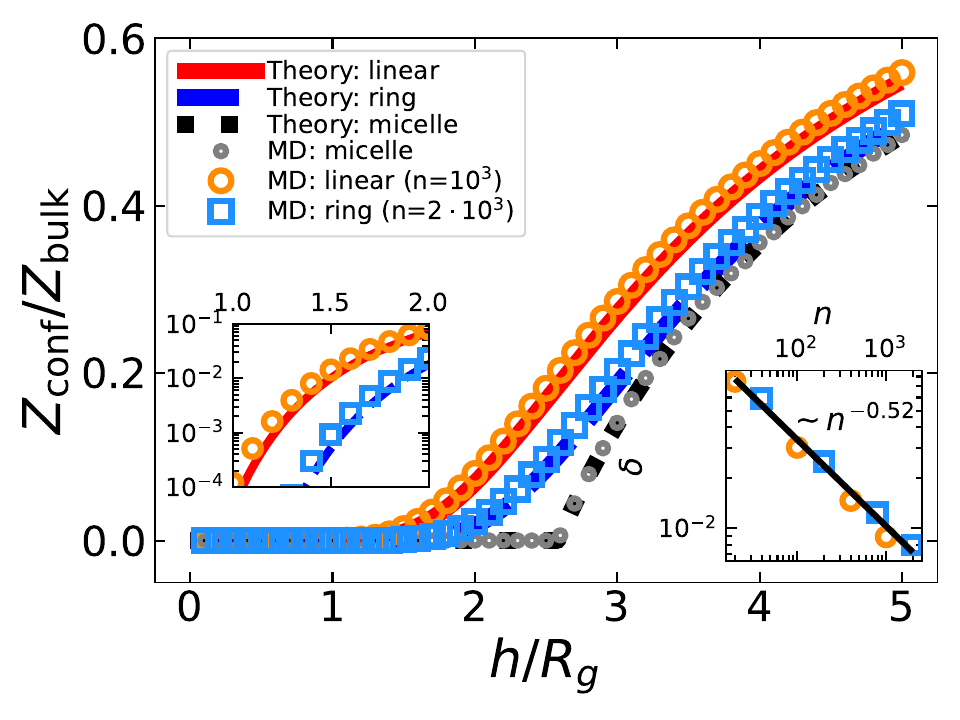}
\caption{}
\label{}
\end{subfigure}
\begin{subfigure}[b]{0.45\textwidth}
\includegraphics[width=1.0\columnwidth]{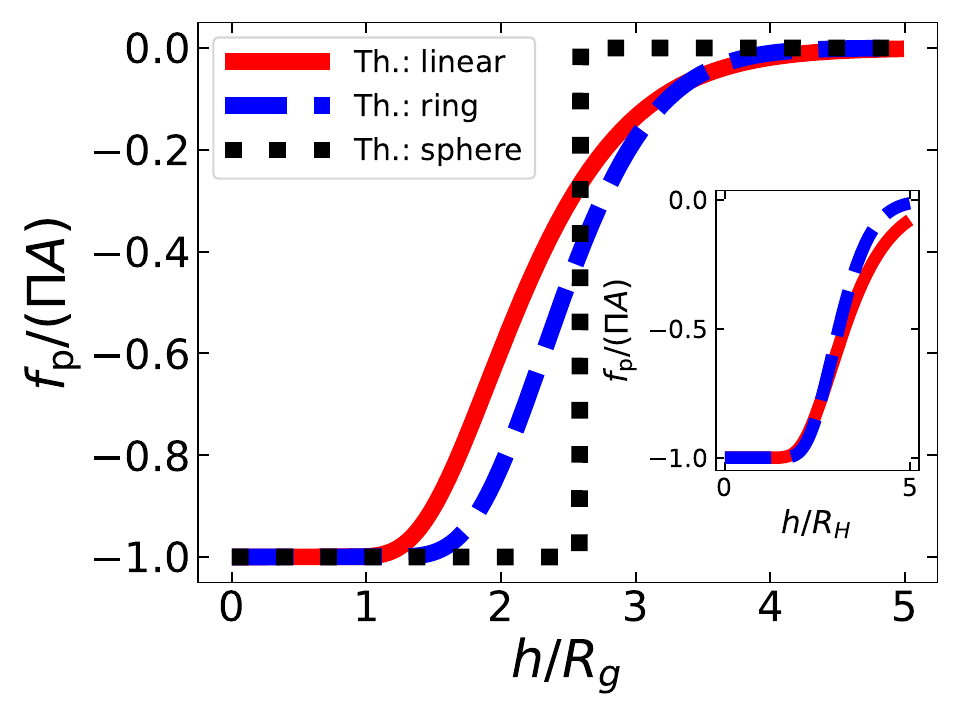}
\caption{}
\label{}
\end{subfigure}
\begin{subfigure}[b]{0.45\textwidth}
\includegraphics[width=1.0\columnwidth]{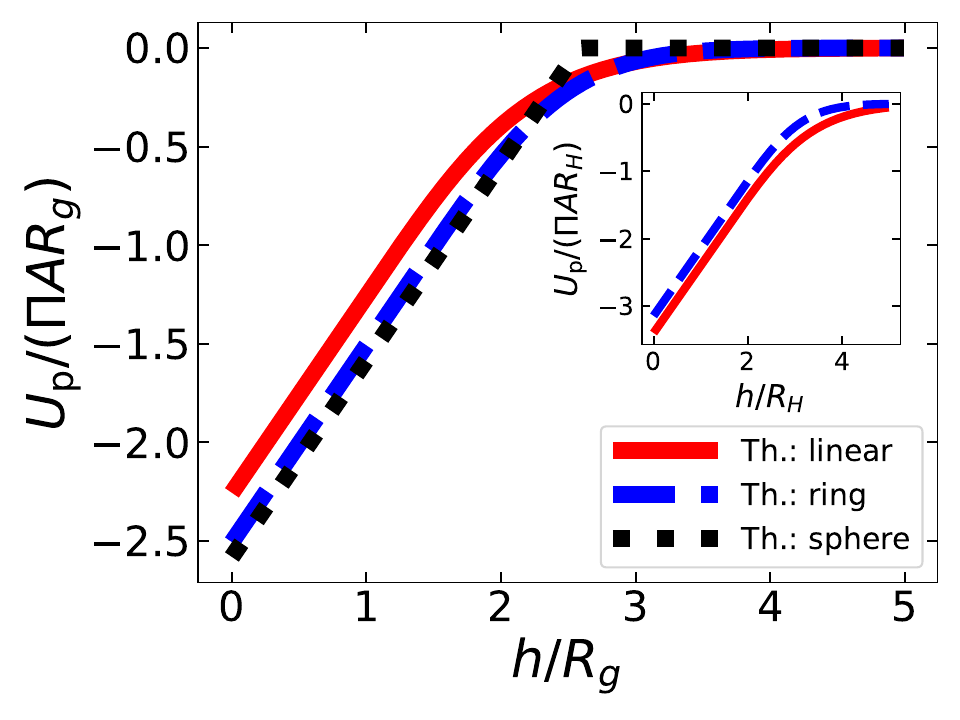}
\caption{}
\label{}
\end{subfigure}
\caption{\label{Fig:Theory+MD} Confinement free energy and depletion interaction between colloidal plates. In all panels, theoretical results for linear polymers, ring polymers, and spherical micelles are shown as  continuous red lines, dashed blue lines, and dotted black lines, respectively. Results from MD simulations are reported as gray dots, orange circles, and light-blue squares for spherical micelles, linear polymers, and ring polymers, respectively. (a) Ratio of the partition function of the polymer under confinement divided by the partition function for the polymer in bulk. The left inset highlights the results for $R_g<h<2R_g$. The inset on the right shows the discrepancy between theory and simulations, $\delta$, as a function of the length of the simulated polymer, where $\delta=\sum_{k=1}^M|y_{MD,i}-y_{Theory}(x_i)|/M$, with $y$ being the ratio of partition functions. The black line indicates a power law fit against all the data using SciPy.~\cite{SciPy} Panels (b) and (c) show the depletion force and the energy caused by various macromolecules. The force is scaled by the osmotic force, $\Pi A$, and the energy is measured in units of osmotic force multiplied by the radius of gyration, $\Pi A R_g$. In the insets of panels (b) and (c) the results are reported with the distance between the plates scaled by the hydrodynamic radius, $R_H$. In the inset of panel (c), the energy is reported relative to $\Pi A R_H$.}
\end{figure*}

The ratio between the confined and bulk partition functions is shown in Fig.~\ref{Fig:Theory+MD}a. 
Polymers can be ``squished'', so in contrast to micelles $Z_\conf/Z_\bulk$ is small but not zero for $h<R_g$. For all depletants $Z_\conf/Z_\bulk\rightarrow 1$ for $h\gg R_g$. Between these two limits $Z_\conf/Z_\bulk$ is larger for linear polymers compared to ring polymers, and it is the smallest for micelles. This suggests that there is a stronger free energetic penalty for moving a ring polymer from bulk to the space between plates, and thus it is more likely to confine linear polymers rather then ring polymers or micelles. 

Figure~\ref{Fig:Theory+MD}b shows the depletion force between the plates induced by the three types of macromolecules. Regardless of the depletant shape, the maximum force is equal to $-\Pi A$, as expected, and for $h > 4 R_g$ the interaction is essentially zero. In the range $1.5 R_g < h < 3 R_g$, the force is always more attractive in the presence of ring polymers as compared to linear polymers, suggesting that indeed they are stronger depletants.
This is confirmed by the potential of depletion interaction between planar colloids in Fig.~\ref{Fig:Theory+MD}c, which clearly shows that the attraction induced by ring polymers is stronger than the attraction induced by linear polymers of equal radii of gyration. 
Overall, the theory suggests that indeed at a given size, measured by radius of gyration, ideal ring polymers are stronger depletants than ideal linear polymers.

Next, we ask whether the result holds if linear and ring polymers have the same hydrodynamic radius, $R_H$, defined to the leading order in $n$ as $R_H^{-1} = N^{-2}\sum_{i\ne j} \langle 1/r_{ij} \rangle$ (see for instance ref.~\cite{Guttman1982Macromolecules}). The hydrodynamic radii for ideal linear and ring polymers with $n\gg1$ are equal to $R_H = (3\sqrt{\pi}/8)R_g$ and $R_H = \sqrt{2/\pi} R_g$, respectively.~\cite{Haydukivska2020SciRep} By replacing $h/R_g$ with $h/R_H\cdot R_H/R_g$ in Eqs.~\ref{Eq:linear}-\ref{Eq:ring}, where $R_H/R_g$ is specific to each depletant, we can write the depletion force and energy exerted by linear and ring polymers of equal hydrodynamic radii. The results in the insets of Fig.~\ref{Fig:Theory+MD}b,c show that in this case it is the linear polymer that is the stronger depletant. 

\begin{figure*}
\begin{subfigure}[b]{0.45\textwidth}
\includegraphics[width=1.0\columnwidth]{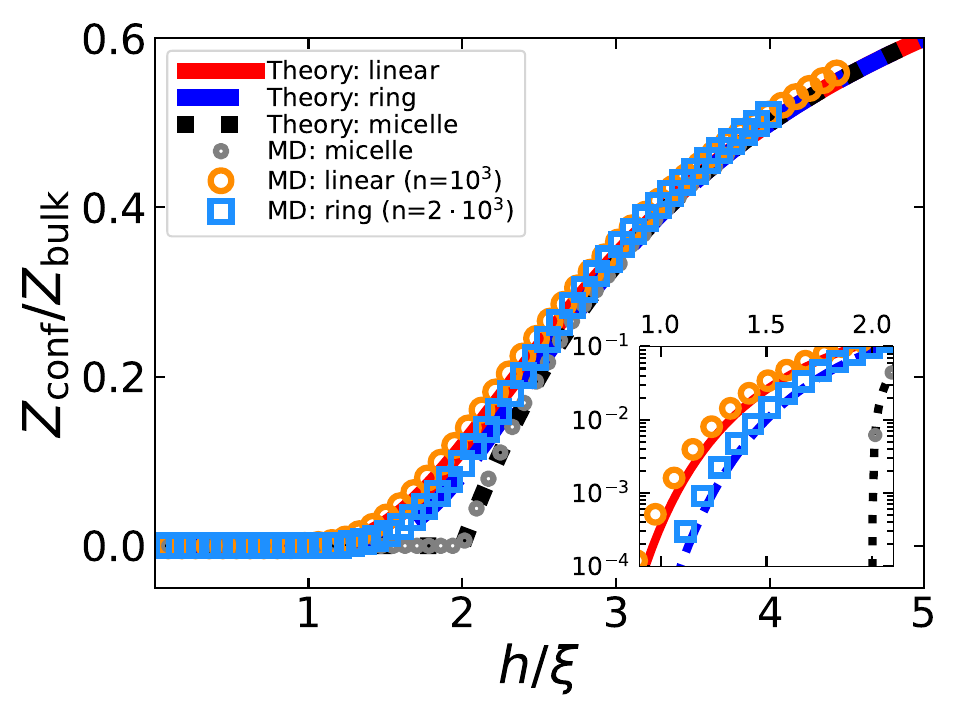}
\caption{}
\label{}
\end{subfigure}
\begin{subfigure}[b]{0.45\textwidth}
\includegraphics[width=1.0\columnwidth]{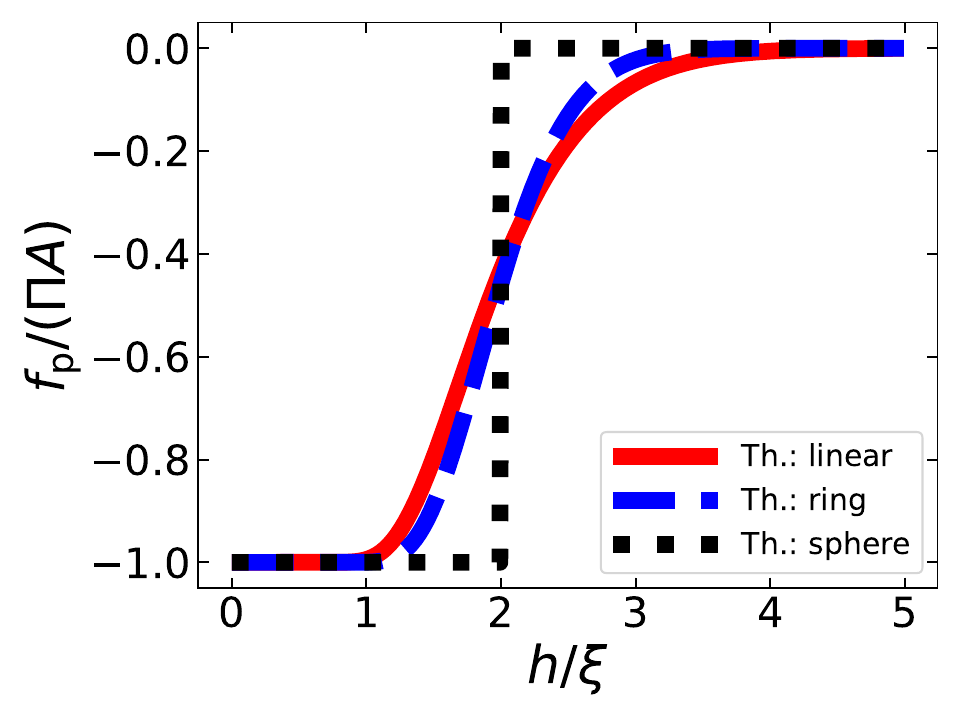}
\caption{}
\label{}
\end{subfigure}
\begin{subfigure}[b]{0.45\textwidth}
\includegraphics[width=1.0\columnwidth]{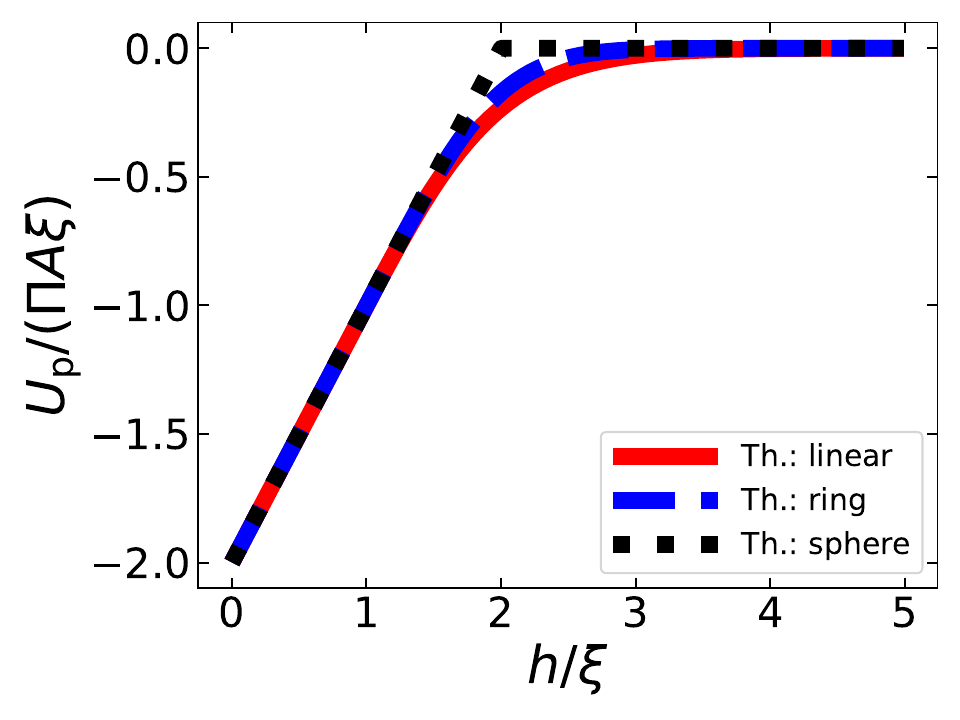}
\caption{}
\label{}
\end{subfigure}
\caption{\label{Fig:Theory+MD+Scale} Scaling by the thickness of the depletion layer, $\xi$. The panels and colors are the same as in Fig.~\ref{Fig:Theory+MD} with the distance between plates scaled by the depletant-specific $\xi$.}
\end{figure*}

Next, we ask whether the differences in depletion interaction between linear and ring polymer can be ascribed to a topology-specific length-scale. We define $\Gamma(h)A$ as the surface excess concentration of the polymer between two plates of area $A$, which is equal to the difference of the total number of polymers found in the chamber in Fig.~\ref{Fig:Model} and the number of polymers in a system of equal volume, temperature, and chemical potential but lacking the colloidal plates (the contribution of other surfaces of the colloids can be ignored because it does not depend on $h$~\cite{Ash1973JChemSocFaradTrans}).  
Tuinier {\it et al}~\cite{Tuinier2000JCP,LekkerkerkerBook} showed that,
\begin{equation}
\Gamma(h) = \rho_{bulk} h\Big[\frac{Z_\conf(hA)}{Z_\bulk(hA)}-1\Big].
\label{Eq:Gamma}
\end{equation}
Therefore, from Eq.~\ref{Eq:fp} we can write the depletion force and potential between colloidal plates as,~\cite{Tuinier2000JCP,LekkerkerkerBook}
\begin{equation}
\begin{aligned}
    f_p(h)&=Ak_BT\frac{\partial}{\partial h}\Gamma(h),\\
    V_p(h)&=\int_h^\infty dh' f_p(h') = Ak_BT[\Gamma(\infty)-\Gamma(h)].
\end{aligned}
\label{Eq:fpGamma}
\end{equation}
$\Gamma(\infty)$ is the surface excess concentration when the colloidal plates are at infinite distance, thus it is equal to twice the surface excess concentration for a single wall~\cite{Tuinier2000JCP,LekkerkerkerBook}. This can be written as $\Gamma(\infty) = 2\int_0^\infty dx [\rho(x)-\rho_{bulk}]$, where $\rho(h)$ is the polymer concentration at distance $h$ from the wall~\cite{Tuinier2000JCP,LekkerkerkerBook,Allain1982PRL}. Clearly $\rho(0)=0$, and for $x\gg R_g$ $\rho(x)\simeq \rho_{bulk}$, so that we can write,~\cite{Tuinier2000JCP,LekkerkerkerBook}
\begin{equation}
\Gamma(\infty)= -2\rho_{bulk}\xi,
\label{Eq:GammaInfty}
\end{equation}
which defines $\xi$ as a measure of a size of the depletion layer.  
Equations~\ref{Eq:Gamma} and Eq.~\ref{Eq:GammaInfty} allow to show that,
\begin{equation}
\xi = -\frac{1}{2}\lim_{h\rightarrow\infty} h\Big[\frac{Z_\conf(hA)}{Z_\bulk(hA)}-1\Big].
\label{Eq:xi}
\end{equation}
Taking the limit in Eq.~\ref{Eq:xi} we obtain $2\xi = 4R_g/\sqrt{\pi}\simeq 2.257 R_g$ for a linear polymer\cite{Gorbunov1984JPolSciUSSR,Tuinier2000JCP,LekkerkerkerBook}
and $2\xi = \sqrt{2\pi}R_g \simeq 2.507 R_g$ for a ring.~\cite{Gorbunov1984JPolSciUSSR}
 For a micelle, $2\xi = 2 \sqrt{5/3} R_g \simeq 2.582 R_g$, that is the diameter of the sphere.
If we now scale the ratio of partition functions, and the force and energy of attraction induced by depletion by $\xi$ (that is we replace $h/R_g$ with $h/\xi \cdot \xi/R_g$ and measure $h$ in units of $\xi$, with $\xi/R_g$ which depends on the depletant), the theoretical results for the ratio of partition functions, the depletion force and the depletion interaction induced by linear and ring polymers become nearly identical, and the energy profile essentially matches the result for spherical, hard micelles. This suggests that this unique length-scale, to a good extent, controls the strength of depletion interaction. Tuinier {\it et al}~\cite{Tuinier2000JCP,LekkerkerkerBook} had already made this observation for linear polymers and micelles. Here, we have shown that the same reasoning can be applied to ring polymers.

\begin{figure}
\begin{subfigure}[b]{0.325\textwidth}
\includegraphics[width=1.0\columnwidth]{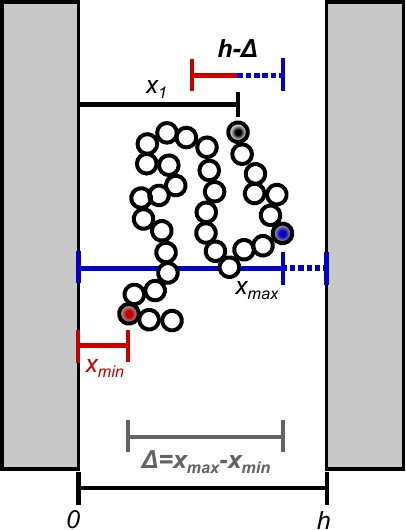}
\caption{}
\label{}
\end{subfigure}
\begin{subfigure}[b]{0.525\textwidth}
\includegraphics[width=1.0\columnwidth]{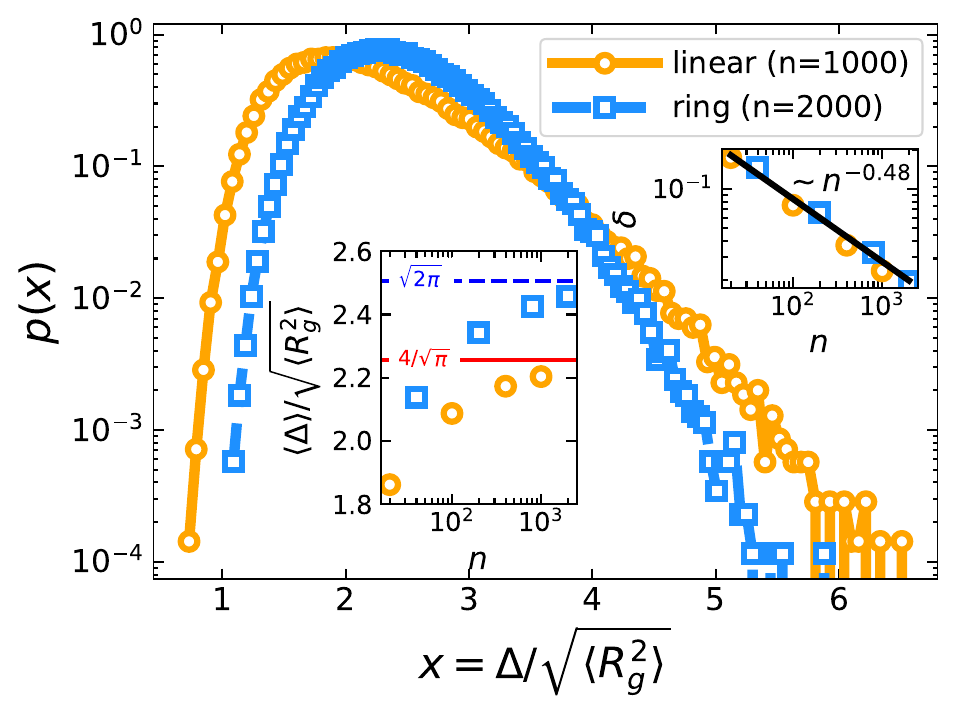}
\caption{}
\label{}
\end{subfigure}
\caption{\label{Fig:Effective} Polymer extension in one direction, $\Delta$. (a) Pictorial representation of $\Delta$. The first bead of the polymer chain ($x_1$) is marked in black, the one closest to the left wall in red ($x_{min}$), and the one nearest to the right wall in blue ($x_{max}$). The distance in the direction perpendicular to the wall between these last two beads is $\Delta$. One can rigidly move the chain by $h-\Delta$ without clashing with the walls. (b) The orange circles and light-blue squares indicate results for linear and ring polymers, respectively. The inset in the middle shows the mean of the distribution as a function of the length of the polymers. The red continuous line and the blue dashed line indicate the theoretical results for linear and ring polymers, respectively. The inset on the right shows the relative error between the theoretical and computational effective radius, $\delta=|\langle \Delta \rangle-2\xi|/(2\xi)$, as a function of polymer length. The black was fit using SciPy~\cite{SciPy} to a power law.
}
\end{figure}

Next, we wonder whether it is possible to obtain a statistical mechanical definition of $\xi$ directly from the coordinates of the depletant. In order to do so, we use the idea behind free energy perturbation~\cite{Zwanzig1954JCP} or particle insertion~\cite{Widom1963JCP}.
We can write,
\begin{equation}
\frac{Z_\conf(h)}{Z_\bulk(h)}=\frac{\int_0^h dx_1\int dx_2\cdots \int dx_N e^{-\beta(U+W)}}{\int_0^h dx_1\int dx_2\cdots \int dx_N e^{-\beta U}}=\langle e^{-\beta W} \rangle,
\label{Eq:ensemble}
\end{equation}
where $W$ is the confinement potential, which is equal to infinity if the $x$ coordinates of any of the particles is $x<0$ or $x>h$ and zero otherwise, and the average $\langle \cdots \rangle$ is performed by sampling all of the conformations of the system in which the first bead of the chain is between $0$ and $h$. We handle this integral using the argument drawn in Fig.~\ref{Fig:Effective}a. First, there are no clashes with the walls if the bead furthest on the right is at $x_{max}<h$ (blue in Fig.~\ref{Fig:Effective}a) and the one furthest on the left is at $x_{min}>0$ (red in Fig.~\ref{Fig:Effective}a). Thus, if we define $\Delta = x_{max}-x_{min}$, that is the maximum extent of the polymer in the direction perpendicular to the plates, we ought to have $h>\Delta$. Moreover, the rigid body translation, in practice the location of the first beads in a fixed conformation, is limited to a segment of length $h-\Delta$ (see Fig.~\ref{Fig:Effective}a). 
Therefore, at the numerator of Eq.~\ref{Eq:ensemble} we can replace the integral in $x_1$ and the wall potential with $(h-\Delta)\Theta(h-\Delta)$, whereas at the denominator no clashes can occur and thus the chain can be rigidly moved between $[0,h]$. This leads us to,
\begin{equation}
\frac{Z_\conf(h)}{Z_\bulk(h)}=\frac{\int dx'_2\cdots \int dx'_N (h-\Delta)\Theta(h-\Delta)e^{-\beta U}}{\int dx'_2\cdots \int dx'_N h e^{-\beta U}}=\Big\langle \Big(1-\frac{\Delta}{h}\Big)\Theta(h-\Delta) \Big\rangle,
\label{Eq:Theory}
\end{equation}
where $x'_i=x_i-x_1$ and the ensemble average $\langle \cdots \rangle$ does not include the rigid body translational degree of freedom of the chain. 
Note that essentially this is the same as Eq.~\ref{Eq:micelle} where the diameter of the sphere has been replaced by $\Delta$ which is a function of the conformation of the polymer.
If $h$ is larger than the maximum value of $\Delta$, we can ignore the step function in Eq.~\ref{Eq:Theory}, and obtain,
\begin{equation}
\frac{Z_\conf(h)}{Z_\bulk(h)} = 1-\frac{\langle\Delta\rangle}{h}\text{;   if  $h > \max(\Delta)$}.
\label{Eq:Theory2}
\end{equation} 
If we plug this result in Eq.~\ref{Eq:xi} we obtain the following identity,
\begin{equation}
2\xi = \langle \Delta \rangle,
\label{Eq:Delta}
\end{equation}
that is the size of the depletion layer $\xi$, which controls the strength of the depletion interaction, is equal to half of the extent of the polymer along one direction averaged over all conformations, $\langle \Delta \rangle$. This intuitive definition allows to compute $\xi$ directly from bulk simulations.

To show this, we perform Molecular Dynamics (MD) simulations of ideal ring and linear polymers using LAMMPS~\cite{LAMMPS}. We consider chains made of beads of mass $m$ which interact only with the previous and next beads in the sequence via a harmonic potential of stiffness $3 k_B T/(2b^2) = 3 \epsilon/(2b^2)$, where $\epsilon=k_BT$ and $b$ are the internal units of energy and length, from which it follows that time has units $\tau = \sqrt{mb^2/\epsilon}$.
The dynamics of the polymer follows the Langevin equation, $m d^2x_i/dt^2 = F_i - m/\eta v_i + F_r$, where $F_i$ is the force due to the interaction with other monomers, $\eta = \tau$ is a drag coefficient, and the random force of zero mean and variance set by the fluctuation-dissipation theorem to be $2 k_BTm/(\eta dt)$, where $dt = 0.02\tau$ is the integration time step.
For simulations of linear polymers of length $n=1000$ beads (ring polymers of length $n=2000$), the total simulations time are $4\times 10^7 \tau$  ($1.8\times10^8 \tau$), and we sampled $n_\mathrm{ens}=4\times10^4$ and ($n_\mathrm{ens}=6\times10^4$) equally spaced conformations.
Shorter simulations were performed to sample conformations of shorter chains.

First, we show that we recover the theoretical results using simulations. We compute $Z_\conf/Z_\bulk$ using Eq.~\ref{Eq:Theory} and average the results over walls perpendicular to $x$, $y$, and $z$. Figures~\ref{Fig:Theory+MD}a and~\ref{Fig:Theory+MD+Scale}a show that the simulations nearly trace the analytical results, although the overlap is not perfect. We reason that the finite size of the chain might cause the discrepancy. Indeed, as shown in the right inset of Fig.~\ref{Fig:Theory+MD}a, the difference between analytical and numerical results decreases approximately as $\sim n^{-1/2}$. This suggests that up to a finite-size effect, the computational model recovers the theoretical results.

As shown in Fig.~\ref{Fig:Effective}b, the distribution of $\Delta$ for linear and ring polymers are both peaked at around $\approx 2\sqrt{\langle R_g^2\rangle}$, though the distribution is narrower for ring polymers. 
The central inset in Fig.~\ref{Fig:Effective} shows that as the number of monomers of the linear and ring polymers increases, $\langle \Delta \rangle$ approaches $2\xi$, and indeed the discrepancy between theory and numerical results roughly decays as $n^{-1/2}$ (see right inset).  
Overall, within finite size effects, MD simulations support the identity between $2\xi$ and $\langle \Delta \rangle$.

In summary, we have shown that (i) at given radius of gyration ideal ring polymers are stronger depletants than ideal linear polymers; (ii) if ideal linear and ring polymers have the same hydrodynamic radius, linear polymers induce a stronger interaction between non-adsorbing colloids; (iii) the difference between ideal linear and ring polymers is largely explained by the size of the depletion layer, $\xi$; (iv) we showed that $2\xi = \langle\Delta\rangle$, that is the extent of the polymer along one direction averaged over all conformations, thus providing a way to compute $\xi$ from bulk simulations.

The experiments by Moghimi {\it et al} suggested that at a given hydrodynamic radius rings are stronger depletants.~\cite{Moghimi2024PhysRevRes} Although we found the opposite we should keep in mind that (i) geometry and size of colloids differ in simulations and experiments; (ii) the theory is valid of polymers in $\Theta$-solvent, whereas experiments are performed in good solvent conditions~\cite{Moghimi2024PhysRevRes}; (iii)  the $\Theta$ temperature depends on polymer topology~\cite{Roovers1983Macromolecules,Lutz1986Makro}; (iv) in $\Theta$-solvent the agreement with experiments of $R_g/R_H$ might not be perfect;~\cite{Guttman1982Macromolecules} (v) finite size effects might be significant.~\cite{Weill1978JdePhys,deGennesBook,Haydukivska2020SciRep}

To conclude, the results presented in this study rationalize the effect of polymer topology on the strength of depletion interaction, and thus can aid the design of depletants which are commonly used to modulate colloidal attraction and self-assembly.~\cite{Sacanna2010Nature,Sananna2015NatMat, Manna2010NanoLett,Moghimi2024PhysRevRes} In addition, our work can be helpful to further characterize the results of experiments involving polymers under confinement, such as the recently introduced escape-time stereometry, which uses the escape time of molecules from entropic traps indented within a microfluidic chip to extract molecular sizes and shapes.~\cite{Zhu2025Science} Finally, as the depletion interaction is expected to be relevant within the cell,~\cite{Merenduzzo2006JCB} our results can also aid the understanding of organization and assembly of cellular structures. 

\begin{acknowledgement}
MLM was supported by the ISMSM-NIST Fellowship at Georgetown University for most of the duration of this project. Simulations were performed on the High-Performance Computing (HPC) systems at Georgetown University. MLM acknowledges the feedback from the Soft Matter community at Georgetown University (in particular Prof. Olmsted, Prof. Blair, Prof. Urbach, Prof. Del Gado, and Dr. Cunha). MLM thanks Dr. Seppala for pointing out a relevant reference, and Dr. Moghimi for numerous conversations about his work as well as this study.
\end{acknowledgement}


\providecommand{\latin}[1]{#1}
\makeatletter
\providecommand{\doi}
  {\begingroup\let\do\@makeother\dospecials
  \catcode`\{=1 \catcode`\}=2 \doi@aux}
\providecommand{\doi@aux}[1]{\endgroup\texttt{#1}}
\makeatother
\providecommand*\mcitethebibliography{\thebibliography}
\csname @ifundefined\endcsname{endmcitethebibliography}
  {\let\endmcitethebibliography\endthebibliography}{}

\end{document}